\documentclass[12 pt,twoside,aps,prd,amsmath,amssymb,
tightenlines,showpacs,showkeys,eqsecnum]{revtex4-1}

\usepackage{graphicx}
\usepackage{mathptmx}
\usepackage{bm}
\newcommand {\ve}{\varepsilon}
\newcommand {\pr}{\partial}
\newcommand {\cG}{\cal G}
\newcommand {\cD}{\cal D}
\newcommand {\bg}{\bar \gamma}
\newcommand {\G}{\Gamma}
\newcommand {\bp}{\bar \psi}
\newcommand {\p}{\psi}

\def\myfigure #1#2#3#4
{\begin{figure}[ht]\begin{center}
\includegraphics[width=#2 \textwidth]{#1.eps}
\parbox[t]{#4\textwidth}{\caption{#3}\label{#1}}
\end{center}\end{figure}}

\def \myfigures #1#2#3#4#5#6#7#8
{\begin{figure}[ht]
    \begin{center}
        \includegraphics[width=#2 \textwidth]{#1.eps}
        \hfill
        \includegraphics[width=#6 \textwidth]{#5.eps}
        \parbox[t]{#4\textwidth}{\caption {#3}\label{#1}}
        \hfill
        \parbox[t]{#8\textwidth}{\caption {#7}\label{#5}}
    \end{center}
\end{figure} }

\begin{document}
\baselineskip -24pt
\title{Nonlinear Spinor Fields in Bianchi type-III spacetime}
\author{Bijan Saha}
\affiliation{Laboratory of Information Technologies\\
Joint Institute for Nuclear Research\\
141980 Dubna, Moscow region, Russia} \email{bijan@jinr.ru}
\homepage{http://spinor.bijansaha.ru}

\begin{abstract}

Within the scope of Bianchi type-III spacetime we study the role of
spinor field on the evolution of the Universe as well as the
influence of gravity on the spinor field. In doing so we have
considered a polynomial type of nonlinearity. In this case the
spacetime remains locally rotationally symmetric and anisotropic all
the time. It is found that depending on the sign of nonlinearity the
models allows both accelerated and oscillatory modes of expansion.
The non-diagonal components of energy-momentum tensor though impose
some restrictions on metric functions and components of spinor
field, unlike Bianchi type I, V and $VI_0$ cases, they do not lead
to vanishing mass and nonlinear terms of the spinor field.
\end{abstract}



\maketitle

\bigskip

\section{Introduction}

Discovery and further reconfirmation of the existence of the late
time accelerated mode of expansion \cite{riess,perlmutter} have
given rise to a number of alternative studies of the evolution of
the Universe.

Though the models with $\Lambda$-term
\cite{starobinsky,PRpadma,2006APSS302-83-91}, quintessence
\cite{Carden,chimento1,Linder1,olivares,zlatev,2005ChineseJPhys43-1035-1043},
Chaplygin gas \cite{Kamenshchik,Amendola,Bean,bento,B1,B2,B4,bilic}
etc. retain their position as the prime candidates to explain this
phenomenon, nevertheless some other models of dark energy are also
proposed. After some remarkable works  by different authors
\cite{henneaux,ochs,saha1997a,saha1997b,saha2001a,greene,saha2004a,
saha2004b,ribas,saha2006c,saha2006e,saha2007,saha2006d,souza,kremer},
showing the important role of spinor field in the evolution of the
Universe, it has been extensively used to model the dark energy.
This success is directly related to its ability to answer some
fundamental questions of modern cosmology: (i) Problem of initial
singularity and its possible elimination
\cite{saha1997a,saha1997b,saha2001a,saha2004a,saha2004b,PopPLB,PopPRD,PopGREG,FabIJTP};
(ii) problem of isotropization
\cite{misner,saha2001a,saha2004a,saha2006c,PopPRD} and (iii) late
time acceleration of the Universe \cite{ribas,
saha2006d,saha2006e,saha2007,PopGREG,PopPLB,FabIJTP,ELKO,FabJMP,PopPRD}.
Moreover recently it was found that the spinor field can also
describe the different characteristics of matter from ekpyrotic
matter to phantom matter, as well as Chaplygin gas
\cite{krechet,saha2010a,saha2010b,saha2011,saha2012}.

In some recent studies \cite{sahaIJTP2014,sahaAPSS2015,sahabvi0} it
was shown that due to its specific behavior in curve spacetime the
spinor field can significantly change not only the geometry of
spacetime but itself as well. The existence of nontrivial
non-diagonal components of the energy-momentum tensor plays a vital
role in this matter. In \cite{sahaIJTP2014,sahaAPSS2015} it was
shown that depending on the type restriction imposed on the
non-diagonal components of the energy-momentum tensor, the initially
Bianchi type-I evolve into a LRS Bianchi type-I spacetime or FRW one
from the very beginning, whereas the model may describe a general
Bianchi type-I spacetime but in that case the spinor field becomes
massless and linear. The same thing happens for a Bianchi
type-$VI_0$ spacetime, i.e., the geometry of Bianchi type-$VI_0$
spacetime does not allow the existence of a massive and nonlinear
spinor field in some particular cases \cite{sahabvi0}. In this paper
we study the role of spinor field on the evolution of a Bianchi
type-III spacetime.

The purpose of considering the anisotropic cosmological models lies
on the fact that recent observational data from Cosmic Background
Explorer's differential Radiometer have detected and measured cosmic
microwave background anisotropies in different angular scale. We
consider here Bianchi type III model due to some special interest of
cosmologist towards it due to its peculiar geometric interpretation
\cite{Coley}. Yadav et al. \cite{YadavMK}, Pradhan et al.
\cite{PAZ,PLA} have recently studied homogeneous and anisotropic
Bianchi type-III spacetime in context of massive strings. Recently
Yadav \cite{YadavYadav} has obtained Bianchi type-III anisotropic DE
models with constant deceleration parameter. In this paper, they
have investigated a new anisotropic Bianchi type-III DE model with
variable $\omega$ without assuming constant deceleration parameter.
A spinor description of dark energy within the scope of a BIII model
was given in \cite{2012IJTP51-1812-1837}. Further the accelerating
dark energy models of the Universe within the scope of Bianchi
type-III spacetime were studied in
\cite{2014EChAYa45-349-396,2015EChAYa46-310-346,Akarsu}. Bianchi
type III cosmological models with varying $\Lambda$ term was studied
in \cite{Adhav}.

\section{Basic equation}

Let us consider the case when the anisotropic spacetime is filled
with nonlinear spinor field. The corresponding action can be given
by
\begin{equation}
{\cal S}(g; \psi, \bp) = \int\, L \sqrt{-g} d\Omega \label{action}
\end{equation}
with
\begin{equation}
L= L_{\rm g} + L_{\rm sp}. \label{lag}
\end{equation}
Here $L_{\rm g}$ corresponds to the gravitational field
\begin{equation}
L_{\rm g} = \frac{R}{2\kappa}, \label{lgrav}
\end{equation}
where $R$ is the scalar curvature, $\kappa = 8 \pi G$, with $G$
being Newton's gravitational constant and $L_{\rm sp}$ is the spinor
field Lagrangian.

\subsection{Gravitational field}

The gravitational field in our case is given by a Bianchi type-III
anisotropic spacetime:

\begin{equation}
ds^2 = dt^2 - a_1^2 e^{-2mx_3} dx_1^2 - a_2^2 dx_2^2 - a_3^2 dx_3^2,
\label{biii}
\end{equation}
with $a_1,\,a_2$ and $a_3$ being the functions of time only and $m$
is some arbitrary constant.

The nontrivial Christoffel symbols for \eqref{biii} are
\begin{eqnarray}
\G_{01}^{1} &=& \frac{\dot{a_1}}{a_1},\quad \G_{02}^{2} =
\frac{\dot{a_2}}{a_2},\quad
\G_{03}^{3} = \frac{\dot{a_3}}{a_3}, \nonumber\\
\G_{11}^{0} &=& a_1 \dot{a_1} e^{-2mx_3},\quad \G_{22}^{0} = a_2
\dot{a_2},\quad
\G_{33}^{0} = a_3 \dot{a_3},\label{Chrysvi}\\
\G_{31}^{1} &=& -m,\quad \G_{11}^{3} = \frac{m a_1^2}{a_3^2}
e^{-2mx_3}. \nonumber
\end{eqnarray}

The nonzero components of the Einstein tensor corresponding to the
metric \eqref{biii} are
\begin{subequations}
\label{ET}
\begin{eqnarray}
G_1^1 &=&  -\frac{\ddot a_2}{a_2} - \frac{\ddot a_3}{a_3} -
\frac{\dot a_2}{a_2}\frac{\dot a_3}{a_3}, \label{ET11}\\
G_2^2 &=&  -\frac{\ddot a_3}{a_3} - \frac{\ddot a_1}{a_1} -
\frac{\dot a_3}{a_3}\frac{\dot a_1}{a_1} + \frac{m^2}{a_3^2}, \label{ET22}\\
G_3^3 &=&  -\frac{\ddot a_1}{a_1} - \frac{\ddot a_2}{a_2} -
\frac{\dot a_1}{a_1}\frac{\dot a_2}{a_2}, \label{ET33}\\
G_0^0 &=&  -\frac{\dot a_1}{a_1}\frac{\dot a_2}{a_2} - \frac{\dot
a_2}{a_2}\frac{\dot a_3}{a_3} - \frac{\dot a_3}{a_3}\frac{\dot
a_1}{a_1} + \frac{m^2}{a_3^2}, \label{ET00}\\
G_3^0 &=& m \left(\frac{\dot a_3}{a_3} -   \frac{\dot
a_1}{a_1}\right). \label{ET03}
\end{eqnarray}
\end{subequations}

\subsection{Spinor field}

For a spinor field $\p$, the symmetry between $\p$ and $\bp$ appears
to demand that one should choose the symmetrized Lagrangian
\cite{kibble}. Keeping this in mind we choose the spinor field
Lagrangian as \cite{saha2001a}:

\begin{equation}
L_{\rm sp} = \frac{\imath}{2} \biggl[\bp \gamma^{\mu} \nabla_{\mu}
\psi- \nabla_{\mu} \bar \psi \gamma^{\mu} \psi \biggr] - m_{\rm sp}
\bp \psi - F, \label{lspin}
\end{equation}
where the nonlinear term $F$ describes the self-interaction of a
spinor field and can be presented as some arbitrary functions of
invariants generated from the real bilinear forms of a spinor field.
Here we consider $F = F(K)$ with $K$ being one of the following
expressions:$\{I,\,J,\,I + J,\,I - J\}$, where $I = S^2 = \left(\bar
\psi \psi\right)^2$ and $J = P^2 = \left(\imath \bar \psi \gamma^5
\psi\right)^2.$ It can be shown that thanks to Fierz identity this
type of nonlinear term describes the nonlinearity in its most
general form. In \eqref{lspin} $\nabla_\mu$ is the covariant
derivative of spinor field:
\begin{equation}
\nabla_\mu \psi = \frac{\partial \psi}{\partial x^\mu} -\G_\mu \psi,
\quad \nabla_\mu \bp = \frac{\partial \bp}{\partial x^\mu} + \bp
\G_\mu, \label{covder}
\end{equation}
with $\G_\mu$ being the spinor affine connection. In \eqref{lspin}
$\gamma$'s are the Dirac matrices in curve spacetime and obey the
following algebra
\begin{equation}
\gamma^\mu \gamma^\nu + \gamma^\nu \gamma^\mu = 2 g^{\mu\nu}
\label{al}
\end{equation}
and are connected with the flat spacetime Dirac matrices $\bg$ in
the following way
\begin{equation}
 g_{\mu \nu} (x)= e_{\mu}^{a}(x) e_{\nu}^{b}(x) \eta_{ab},
\quad \gamma_\mu(x)= e_{\mu}^{a}(x) \bg_a \label{dg}
\end{equation}
where $e_{\mu}^{a}$ is a set of tetrad 4-vectors.

For the metric \eqref{biii} we choose the tetrad as follows:

\begin{equation}
e_0^{(0)} = 1, \quad e_1^{(1)} = a_1 e^{-mx_3}, \quad e_2^{(2)} =
a_2, \quad e_3^{(3)} = a_3. \label{tetradvi}
\end{equation}

The Dirac matrices $\gamma^\mu(x)$ of Bianchi type-III spacetime are
connected with those of Minkowski one as follows:
$$ \gamma^0=\bg^0,\quad \gamma^1 = \frac{ e^{m x_3}}{a_1} \bg^1,
\quad \gamma^2= \frac{1}{a_2}\bg^2,\quad \gamma^3 = \frac{
1}{a_3}\bg^3$$

$$\gamma^5 = - \imath \sqrt{-g}
\gamma^0\gamma^1\gamma^2\gamma^3 = - \imath \bg^0\bg^1\bg^2\bg^3 =
\bg^5
$$
with
\begin{eqnarray}
\bg^0\,=\,\left(\begin{array}{cc}I&0\\0&-I\end{array}\right), \quad
\bg^i\,=\,\left(\begin{array}{cc}0&\sigma^i\\
-\sigma^i&0\end{array}\right), \quad
\gamma^5 = \bg^5&=&\left(\begin{array}{cc}0&-I\\
-I&0\end{array}\right),\nonumber
\end{eqnarray}
where $\sigma_i$ are the Pauli matrices:
\begin{eqnarray}
\sigma^1\,=\,\left(\begin{array}{cc}0&1\\1&0\end{array}\right),
\quad \sigma^2\,=\,\left(\begin{array}{cc}0&-\imath\\
\imath&0\end{array}\right), \quad
\sigma^3\,=\,\left(\begin{array}{cc}1&0\\0&-1\end{array}\right).
\nonumber
\end{eqnarray}
Note that the $\bg$ and the $\sigma$ matrices obey the following
properties:
\begin{eqnarray}
\bg^i \bg^j + \bg^j \bg^i = 2 \eta^{ij},\quad i,j = 0,1,2,3
\nonumber\\
\bg^i \bg^5 + \bg^5 \bg^i = 0, \quad (\bg^5)^2 = I,
\quad i=0,1,2,3 \nonumber\\
\sigma^j \sigma^k = \delta_{jk} + i \varepsilon_{jkl} \sigma^l,
\quad j,k,l = 1,2,3 \nonumber
\end{eqnarray}
where $\eta_{ij} = \{1,-1,-1,-1\}$ is the diagonal matrix,
$\delta_{jk}$ is the Kronekar symbol and $\varepsilon_{jkl}$ is the
totally antisymmetric matrix with $\varepsilon_{123} = +1$.

The spinor affine connection matrices $\G_\mu (x)$ are uniquely
determined up to an additive multiple of the unit matrix by the
equation
\begin{equation}
\frac{\pr \gamma_\nu}{\pr x^\mu} - \G_{\nu\mu}^{\rho}\gamma_\rho -
\G_\mu \gamma_\nu + \gamma_\nu \G_\mu = 0, \label{afsp}
\end{equation}
with the solution
\begin{equation}
\Gamma_\mu = \frac{1}{4} \bg_{a} \gamma^\nu \partial_\mu e^{(a)}_\nu
- \frac{1}{4} \gamma_\rho \gamma^\nu \Gamma^{\rho}_{\mu\nu}.
\label{sfc}
\end{equation}

From the Bianchi type-VI metric \eqref{sfc} one finds the following
expressions for spinor affine connections:
\begin{subequations}
\label{sac123}
\begin{eqnarray}
\G_0 &=& 0, \label{sac0}\\  \G_1 &=& \frac{1}{2}\Bigl(\dot a_1
\bg^1\bg^0 - m\frac{a_1}{a_3} \bg^1\bg^3\Bigr) e^{-mx_3},
\label{sac1}\\  \G_2 &=& \frac{1}{2}\dot a_2 \bg^2\bg^0,
\label{sac2}\\  \G_3 &=& \frac{1}{2} \dot a_3 \bg^3 \bg^0.
\label{sac3}
\end{eqnarray}
\end{subequations}

\subsection{Field equations}

Variation of \eqref{action} with respect to the metric function
$g_{\mu \nu}$ gives the Einstein field equation
\begin{equation}
G_\mu^\nu = R_\mu^\nu - \frac{1}{2} \delta_\mu^\nu R = -\kappa
T_\mu^\nu, \label{EEg}
\end{equation}
where $R_\mu^\nu$ and $R$ are the Ricci tensor and Ricci scalar,
respectively. Here $T_\mu^\nu$ is the energy momentum tensor of the
spinor field.

Varying \eqref{lspin} with respect to $\bp (\psi)$ one finds the
spinor field equations:
\begin{subequations}
\label{speq}
\begin{eqnarray}
\imath\gamma^\mu \nabla_\mu \psi - m_{\rm sp} \psi - {\cD} \psi -
 \imath {\cG} \gamma^5 \psi &=&0, \label{speq1} \\
\imath \nabla_\mu \bp \gamma^\mu +  m_{\rm sp} \bp + {\cD}\bp +
\imath {\cG} \bp \gamma^5 &=& 0, \label{speq2}
\end{eqnarray}
\end{subequations}
where we denote ${\cD} = 2 S F_K K_I$ and ${\cG} = 2 P F_K K_J$,
with $F_K = dF/dK$, $K_I = dK/dI$ and $K_J = dK/dJ.$ In view of
\eqref{speq} the spinor field Lagrangian \eqref{lspin}  can be
rewritten as
\begin{eqnarray}
L_{\rm sp} & = & \frac{\imath}{2} \bigl[\bp \gamma^{\mu}
\nabla_{\mu} \psi- \nabla_{\mu} \bar \psi \gamma^{\mu} \psi \bigr] -
m_{\rm sp} \bp \psi - F(I,\,J)
\nonumber \\
& = & \frac{\imath}{2} \bp [\gamma^{\mu} \nabla_{\mu} \psi - m_{\rm
sp} \psi] - \frac{\imath}{2}[\nabla_{\mu} \bar \psi \gamma^{\mu} +
m_{\rm sp} \bp] \psi
- F(I,\,J),\nonumber \\
& = & 2 (I F_I + J F_J) - F = 2 K F_K - F(K). \label{lspin01}
\end{eqnarray}

\subsection{Energy momentum tensor of the spinor field}

The energy-momentum tensor of the spinor field is given by
\begin{equation}
T_{\mu}^{\rho}=\frac{\imath}{4} g^{\rho\nu} \biggl(\bp \gamma_\mu
\nabla_\nu \psi + \bp \gamma_\nu \nabla_\mu \psi - \nabla_\mu \bar
\psi \gamma_\nu \psi - \nabla_\nu \bp \gamma_\mu \psi \biggr) \,-
\delta_{\mu}^{\rho} L_{\rm sp}. \label{temsp}
\end{equation}

Then in view of \eqref{covder} and \eqref{lspin01} the
energy-momentum tensor of the spinor field can be written as
\begin{eqnarray}
T_{\mu}^{\,\,\,\rho}&=&\frac{\imath}{4} g^{\rho\nu} \bigl(\bp
\gamma_\mu
\partial_\nu \psi + \bp \gamma_\nu \partial_\mu \psi -
\partial_\mu \bar \psi \gamma_\nu \psi - \partial_\nu \bp \gamma_\mu
\psi \bigr)\nonumber\\
& - &\frac{\imath}{4} g^{\rho\nu} \bp \bigl(\gamma_\mu \G_\nu +
\G_\nu \gamma_\mu + \gamma_\nu \G_\mu + \G_\mu \gamma_\nu\bigr)\psi
 \,- \delta_{\mu}^{\rho} \bigl(2 K F_K - F(K)\bigr). \label{temsp0}
\end{eqnarray}
As is seen from \eqref{temsp0}, is case if for a given metric
$\G_\mu$'s are different, there arise nontrivial non-diagonal
components of the energy momentum tensor.

We consider the case when the spinor field depends on $t$ only. Then
after a little manipulations from \eqref{temsp0} for the components
of the energy momentum tensor one finds:
\begin{subequations}
\label{Ttot}
\begin{eqnarray}
T_0^0 & = & m_{\rm sp} S + F(K), \label{emt00}\\
T_1^1 &=& T_2^2 = T_3^3 =  F(K) - 2 K F_K, \label{emtii}\\
T_1^0 &=& 0, \label{emt01} \\
T_2^0 &=&-\frac{\imath}{4} m \frac{a_2}{a_3}\, \bp \bg^2 \bg^3 \bg^0
\psi
= -\frac{1}{4} m \frac{a_2}{a_3} \,A^1, \label{emt02} \\
T_3^0 &=& 0, \label{emt03} \\
T_2^1 &=& \frac{\imath}{4} \frac{a_2}{a_1} e^{m x_3}
\biggl[\biggl(\frac{\dot a_1}{a_1} - \frac{\dot a_2}{a_2}\biggr) \bp
\bg^1 \bg^2 \bg^0 \psi - \frac{m}{a_3} \bp \bg^1 \bg^2 \bg^3 \psi
\biggr] \nonumber \\
&=& \frac{1}{4} \frac{a_2}{a_1} e^{m x_3} \biggl[\biggl(\frac{\dot
a_1}{a_1} - \frac{\dot a_2}{a_2}\biggr) A^3
- \frac{m}{a_3}A^0\biggr] , \label{emt12}\\
T_3^1 &=&\frac{\imath}{4} \frac{a_3}{a_1} e^{m x_3}
\biggl(\frac{\dot a_3}{a_3} - \frac{\dot a_1}{a_1}\biggr) \bp \bg^3
\bg^1 \bg^0 \psi = \frac{1}{4} \frac{a_3}{a_1} e^{m x_3}
\biggl(\frac{\dot a_3}{a_3} - \frac{\dot a_1}{a_1}\biggr) A^2 \label{emt13}\\
T_3^2 &=&\frac{\imath}{4} \frac{a_3}{a_2} \biggl(\frac{\dot
a_2}{a_2} - \frac{\dot a_3}{a_3}\biggr) \bp \bg^2 \bg^3 \bg^0 \psi =
\frac{1}{4} \frac{a_3}{a_2} \biggl(\frac{\dot a_2}{a_2} - \frac{\dot
a_3}{a_3}\biggr)A^1. \label{emt23}
\end{eqnarray}
\end{subequations}

It can be shown that bilinear spinor forms the obey the following
system of equations:
\begin{subequations}
\label{inv}
\begin{eqnarray}
\dot S_0  +  {\cG} A_{0}^{0} &=& 0, \label{S0} \\
\dot P_0  -  \Phi A_{0}^{0} &=& 0, \label{P0}\\
\dot A_{0}^{0} -\frac{m}{a_3} A_{0}^{3} +  \Phi P_0 -  {\cG}
S_0 &=& 0, \label{A00}\\
\dot A_{0}^{3} -\frac{m}{a_3} A_{0}^{0} &=& 0, \label{A03}\\
\dot v_{0}^{0} - \frac{m}{a_3} v_{0}^{3} &=& 0,\label{v00} \\
\dot v_{0}^{3} - \frac{m}{a_3} v_{0}^{0} +
\Phi Q_{0}^{30} +  {\cG} Q_{0}^{21} &=& 0,\label{v03}\\
\dot Q_{0}^{30}  -  \Phi v_{0}^{3} &=& 0,\label{Q030} \\
\dot Q_{0}^{21}  -  {\cG} v_{0}^{3} &=& 0, \label{Q021}
\end{eqnarray}
\end{subequations}
where we denote where we denote $S_0 = S V,\, P_0 = P V,\, A_0^\mu =
A_\mu V,\, v_0^\mu = v^\mu V,\, Q_0^{\mu \nu} = Q^{\mu \nu} V$ and
$\Phi = m_{\rm sp} + {\cD}$. Here
 $S = \bar \psi \psi$ is a scalar, $
 P =  \imath \bar \psi \gamma^5 \psi$ is a pseudoscalar,
$ v^\mu = \bar \psi \gamma^\mu \psi$ - vector, $ A^\mu = \bar \psi
\gamma^5 \gamma^\mu \psi$ - pseudovector, and  $Q^{\mu\nu} =\bar
\psi \sigma^{\mu\nu} \psi$ is antisymmetric tensor. In \eqref{inv}
$V$ is the volume scale which is defined as
\begin{equation}
V = a_1 a_2 a_3. \label{VDef}
\end{equation}

Combining these equations together and taking the first integral one
gets
\begin{subequations}
\label{inv0}
\begin{eqnarray}
(S_{0})^{2} + (P_{0})^{2} + (A_{0}^{0})^{2} - (A_{0}^{3})^{2} &=&
C_1 = {\rm Const}, \label{inv01}\\
(Q_{0}^{30})^{2} + (Q_{0}^{21})^{2} + (v_{0}^{3})^{2} -
(v_{0}^{0})^{2} &=& C_2 = {\rm Const} \label{inv02}
\end{eqnarray}
\end{subequations}

Now let us consider the Einstein field equations. In view of
\eqref{ET} and \eqref{Ttot} with find the following system of
Einstein Equations

\begin{subequations}
\label{EE}
\begin{eqnarray}
\frac{\ddot a_2}{a_2} + \frac{\ddot a_3}{a_3} +
\frac{\dot a_2}{a_2}\frac{\dot a_3}{a_3} &=& \kappa\bigl(F(K) - 2 K F_K\bigr), \label{EE11}\\
\frac{\ddot a_3}{a_3} + \frac{\ddot a_1}{a_1} +
\frac{\dot a_3}{a_3}\frac{\dot a_1}{a_1} - \frac{m^2}{a_3^2} &=& \kappa\bigl(F(K) - 2 K F_K\bigr), \label{EE22}\\
\frac{\ddot a_1}{a_1} + \frac{\ddot a_2}{a_2} +
\frac{\dot a_1}{a_1}\frac{\dot a_2}{a_2} &=& \kappa\bigl(F(K) - 2 K F_K\bigr), \label{EE33}\\
\frac{\dot a_1}{a_1}\frac{\dot a_2}{a_2} + \frac{\dot
a_2}{a_2}\frac{\dot a_3}{a_3} + \frac{\dot a_3}{a_3}\frac{\dot
a_1}{a_1} - \frac{m^2}{a_3^2} &=&   \kappa\bigl(m_{\rm sp} S +
F(K)\bigr), \label{EE00}\\
m \bigl(\frac{\dot a_3}{a_3} - \frac{\dot a_1}{a_1}\bigr) &=& 0,
\label{EE03}\\
0  &=& \frac{1}{4} m \frac{a_2}{a_3}\,A^1, \label{AC02} \\
0  &=& \frac{1}{4} \frac{a_2}{a_1} e^{m x_3}
\biggl[\biggl(\frac{\dot a_1}{a_1} - \frac{\dot a_2}{a_2}\biggr) A^3
- \frac{m }{a_3}A^0\biggr], \label{AC12}\\
0  &=& \frac{1}{4} \frac{a_3}{a_1} e^{m x_3}
\biggl(\frac{\dot a_3}{a_3} - \frac{\dot a_1}{a_1}\biggr) A^2, \label{AC13}\\
0  &=&\frac{1}{4} \frac{a_3}{a_2} \biggl(\frac{\dot a_2}{a_2} -
\frac{\dot a_3}{a_3}\biggr)A^1. \label{AC23}
\end{eqnarray}
\end{subequations}

From \eqref{AC02} one immediately finds
\begin{equation}
 A^1 = 0, \label{A12}
\end{equation}

whereas from \eqref{EE03} one finds the following relation between
$a_1$ and $a_3$:
\begin{equation}
a_3 = X_0 a_1, \quad X_0 = {\rm const.} \label{abcrel}
\end{equation}

In view of \eqref{A12} the relations  \eqref{AC23} fulfill even
without imposing restrictions on the metric functions, whereas
\eqref{AC13} fulfills thanks to \eqref{EE03}. From \eqref{AC12} one
finds the following relations between $A^0$ and $A^3$:
\begin{equation}
\biggl(\frac{\dot a_1}{a_1} - \frac{\dot a_2}{a_2}\biggr) A^3 =
\frac{m }{a_3}A^0. \label{A0A3}
\end{equation}
Inserting \eqref{A0A3} into \eqref{A03} one finds
\begin{equation}
\frac{\dot A_{0}^{3}}{A_{0}^{3}} = \biggl(\frac{\dot a_1}{a_1} -
\frac{\dot a_2}{a_2}\biggr), \label{A3a12}
\end{equation}
with the solution
\begin{equation}
A_{0}^{3} = X_{03} \bigl(\frac{ a_1}{a_2}\bigr), \quad X_{03} = {\rm
const.} \label{A03a12}
\end{equation}

As it was found in previous papers, due to explicit presence of
$a_3$ in the Einstein equations, one needs some additional
conditions. In an early work we propose two different situation,
namely, set $a_3 = \sqrt{V}$ and $a_3 = V$ which allows us to obtain
exact solutions for the metric functions.

In a recent paper we imposed the proportionality condition, widely
used in literature. Demanding that the expansion is proportion to a
component of the shear tensor, namely
\begin{equation}
\vartheta = N_3 \sigma_3^3, \quad N_3 = {\rm const.}
\label{propconiii}
\end{equation}
The motivation behind assuming this condition is explained with
reference to  Thorne \cite{thorne67}, the observations of the
velocity-red-shift relation for extragalactic sources suggest that
Hubble expansion of the universe is isotropic today within $\approx
30$ per cent \cite{kans66,ks66}. To put more precisely, red-shift
studies place the limit
\begin{equation}
\frac{\sigma}{H} \leq 0.3, \label{propconiiiexp}
\end{equation}
on the ratio of shear $\sigma$ to Hubble constant $H$ in the
neighborhood of our Galaxy today. Collins et al. \cite{Collins} have
pointed out that for spatially homogeneous metric, the normal
congruence to the homogeneous expansion satisfies that the condition
$\frac{\sigma}{\theta}$ is constant. Under this proportionality
condition it was also found that the energy-momentum distribution of
the model is strictly isotropic, which is absolutely true for our
case.

In order to exploit the proportionality condition \eqref{propconiii}
Let us now find expansion and shear for BIII metric. The expansion
is given by
\begin{equation}
\vartheta = u^\mu_{;\mu} = u^\mu_{\mu} + \G^\mu_{\mu\alpha}
u^\alpha, \label{expansion}
\end{equation}
and the shear is given by
\begin{equation}
\sigma^2 = \frac{1}{2} \sigma_{\mu\nu} \sigma^{\mu\nu},
\label{shear}
\end{equation}
with
\begin{equation}
\sigma_{\mu\nu} = \frac{1}{2}\bigl[u_{\mu;\alpha} P^\alpha_\nu +
u_{\nu;\alpha} P^\alpha_\mu \bigr] - \frac{1}{3} \vartheta
P_{\mu\nu}, \label{shearcomp}
\end{equation}
where the projection vector $P$:
\begin{equation}
P^2 = P, \quad P_{\mu\nu} = g_{\mu\nu} - u_\mu u_\nu, \quad
P^\mu_\nu = \delta^\mu_\nu - u^\mu u_\nu. \label{proj}
\end{equation}
In comoving system we have $u^\mu = (1,0,0,0)$. In this case one
finds
\begin{equation}
\vartheta = \frac{\dot a_1}{a_1} + \frac{\dot a_2}{a_2} + \frac{\dot
a_3}{a_3} = \frac{\dot V}{V}, \label{expbiii}
\end{equation}
and
\begin{subequations}
\label{shearcomps}
\begin{eqnarray}
\sigma_{1}^{1} &=& -\frac{1}{3}\Bigl(-2\frac{\dot a_1}{a_1} +
\frac{\dot
a_2}{a_2} + \frac{\dot a_3}{a_3}\Bigr) =  \frac{\dot a_1}{a_1} - \frac{1}{3} \vartheta, \label{sh11}\\
\sigma_{2}^{2} &=& -\frac{1}{3}\Bigl(-2\frac{\dot a_2}{a_2} +
\frac{\dot a_3}{a_3} +
\frac{\dot a_1}{a_1}\Bigr) =  \frac{\dot a_2}{a_2} - \frac{1}{3} \vartheta, \label{sh22}\\
\sigma_{3}^{3} &=& -\frac{1}{3}\Bigl(-2\frac{\dot a_3}{a_3} +
\frac{\dot a_1}{a_1} + \frac{\dot a_2}{a_2}\Bigr) =  \frac{\dot
a_3}{a_3} - \frac{1}{3} \vartheta. \label{sh33}
\end{eqnarray}
\end{subequations}

One then finds
\begin{equation}
\sigma^ 2 = \frac{1}{2}\biggl[\sum_{i=1}^3 \biggl(\frac{\dot
a_i}{a_i}\biggr)^2 - \frac{1}{3}\vartheta^2\biggr] =
\frac{1}{2}\biggl[\sum_{i=1}^3 H_i^2 -
\frac{1}{3}\vartheta^2\biggr]. \label{sheargen}
\end{equation}

Inserting \eqref{abcrel}  into \eqref{expbiii}, \eqref{shearcomps}
and \eqref{sheargen} we find

\begin{equation}
\vartheta = 2 \frac{\dot a_1}{a_1} + \frac{\dot a_2}{a_2},
\label{expbvi1}
\end{equation}
and
\begin{subequations}
\label{shearcomps0}
\begin{eqnarray}
\sigma_{1}^{1} &=& \frac{1}{3}\Bigl(\frac{\dot a_1}{a_1} -
\frac{\dot a_2}{a_2}\Bigr), \label{sh110}\\
\sigma_{2}^{2} &=&  -\frac{2}{3}\Bigl(\frac{\dot a_1}{a_1}
- \frac{\dot a_2}{a_2}\Bigr), \label{sh220}\\
\sigma_{3}^{3} &=& \frac{1}{3}\Bigl(\frac{\dot a_1}{a_1} -
\frac{\dot a_2}{a_2}\Bigr). \label{sh330}
\end{eqnarray}
\end{subequations}

On account of \eqref{abcrel}, \eqref{sh33}, \eqref{VDef} from
\eqref{propconiii} one finds
\begin{eqnarray}
a_1 = \Biggl[\frac{1}{X_0 X_1}\,V\Biggr]^{\frac{1}{3} + \frac{1
}{N_3}}, \quad a_2 = X_1 \Biggl[\frac{1}{X_0
X_1}\,V\Biggr]^{\frac{1}{3} - \frac{2}{N_3}},\quad a_3 = X_0
\Biggl[\frac{1}{X_0 X_1}\,V\Biggr]^{\frac{1}{3} + \frac{1}{N_3}},
\label{Metf}
\end{eqnarray}
where $X_1$ is an integration constant. As one sees from
\eqref{Metf}, the isotropization process can take place only for
$N_3 \gg 3$.

The equation for $V$ can be found from the Einstein Equation
\eqref{ET} which for some manipulation looks
\begin{equation}
\ddot V = {\bar X} V^{\frac{1}{3} - \frac{2}{N_3}} + \frac{3
\kappa}{2} \bigl[m_{\rm sp} S + 2 \bigl(F(K) - K F_K\bigr)\bigr] V,
\quad {\bar X} = 2 m^2 X_0^{\frac{2}{N_3} - \frac{4}{3}}
X_1^{\frac{2}{N_3} + \frac{2}{3}}. \label{Vdefein}
\end{equation}
In order to solve \eqref{Vdefein} we have to know the relation
between the spinor and the gravitational fields. Let us first find
those relations for different $K$.

In case of $K = I$, i.e. ${\cG} = 0$ from \eqref{S0} we duly have
\begin{equation}
\dot S_0 = 0, \label{S0n}
\end{equation}
with the solution
\begin{equation}
S = \frac{V_0}{V}, \quad \Rightarrow \quad K = I = S^2 =
\frac{V_0^2}{V^2},   \quad V_0 = {\rm const.} \label{SV}
\end{equation}
In  this case spinor field can be either massive or massless.

As far as case with $K = \{J,\,I+J,\,I-J\}$ that gives $K_J = \pm 1$
is concerned, it can be solved exactly only for a massless spinor
field.

In case of $K = J$, i.e. $ \Phi = {\cD} = 0$ from \eqref{P0} we duly
have
\begin{equation}
\dot P_0 = 0, \label{P0n}
\end{equation}
with the solution
\begin{equation}
P = \frac{V_0}{V}, \quad \Rightarrow \quad  K = J = P^2 =
\frac{V_0^2}{V^2},  \quad V_0 = {\rm const.} \label{PV}
\end{equation}

In case of $K = I + J$ the equations \eqref{S0} and \eqref{P0} can
be rewritten as
\begin{subequations}
\begin{eqnarray}
\dot S_0  +  2 P F_K  A_{0}^{0} &=& 0, \label{S0new} \\
\dot P_0  -  2 S F_K  A_{0}^{0} &=& 0, \label{P0new}
\end{eqnarray}
\end{subequations}
which can be rearranged as
\begin{equation}
S_0 \dot S_0 +  P_0 \dot P_0 = \frac{d}{dt}\bigl( S_0^2 +
P_0^2\bigr) = \frac{d}{dt}\bigl(V^2 K\bigr) = 0, \label{K0}
\end{equation}
with the solution
\begin{equation}
K = I + J = \frac{V_0^2}{V^2}, \quad V_0 = {\rm const.} \label{KV}
\end{equation}
Note that one can represent $S$ and $P$ as follows:
\begin{equation}
S =  \frac{V_0}{V} \sin{\theta}, \quad P = \frac{V_0}{V}
\cos{\theta}. \label{KIpJ}
\end{equation}
The term $\theta$ can be determined from \eqref{S0new} or
\eqref{P0new} on account of \eqref{A0A3}, \eqref{A03a12} and
\eqref{Metf}. It can be shown that $\theta = \theta (V)$.

Finally, for $K = I - J$ the equations \eqref{S0} and \eqref{P0} can
be rewritten as
\begin{subequations}
\begin{eqnarray}
\dot S_0  -  2 P F_K  A_{0}^{0} &=& 0, \label{S0new1} \\
\dot P_0  -  2 S F_K  A_{0}^{0} &=& 0, \label{P0new1}
\end{eqnarray}
\end{subequations}
which can be rearranged as
\begin{equation}
S_0 \dot S_0 -  P_0 \dot P_0 = \frac{d}{dt}\bigl( S_0^2 -
P_0^2\bigr) = \frac{d}{dt}\bigl(V^2 K\bigr) = 0, \label{K01}
\end{equation}
with the solution
\begin{equation}
K = I - J = \frac{V_0^2}{V^2}, \quad V_0 = {\rm const.} \label{KV1}
\end{equation}
As in previous case one can rewrite $S$ and $P$ as follows:
\begin{equation}
S = \frac{V_0}{V} \cosh{\theta}, \quad P = \frac{V_0}{V}
\sinh{\theta}. \label{KImJ}
\end{equation}

As one sees the non-triviality of non-diagonal components of the
energy-momentum tensors, namely $T_2^1$, $T_3^2$ and $T_3^1$ is
directly connected with the affine spinor connections $\G_i$'s.

\section{Solution to the field equations}

In this section we solve the field equations. Let us begin with the
spinor field equations. In view of \eqref{covder} and \eqref{sac123}
the spinor field equation \eqref{speq1} takes the form

\begin{subequations}
\label{SF1}
\begin{eqnarray}
\imath \bg^0 \bigl(\dot \psi + \frac{1}{2}\frac{\dot V}{V}
\psi\bigr) - m_{\rm sp} \psi -\frac{m}{2 a_3} \bg^3 \psi - {\cD}
\psi -  \imath {\cG} \bg^5 \psi &=&0, \label{speq1p}\\
\imath \bigl(\dot \bp + \frac{1}{2}\frac{\dot V}{V} \bp\bigr)\bg^0 +
m_{\rm sp} \bp -\frac{m}{2 a_3}\bp \bg^3   + {\cD}  \bp + \imath
{\cG}\bp \bg^5 &=& 0. \label{speq2p}
\end{eqnarray}
\end{subequations}

As we have already mentioned, $\psi$ is a function of $t$ only. We
consider the 4-component spinor field given by
\begin{eqnarray}
\psi = \left(\begin{array}{c} \psi_1\\ \psi_2\\ \psi_3 \\
\psi_4\end{array}\right). \label{psi}
\end{eqnarray}
Denoting $\phi_i =\sqrt{V} \psi_i$ and $\bar X_0 =  m X_0^{1/N_3 -
2/3} X_1^{1/N_3 + 1/3}$ from \eqref{SF1} for the spinor field we
find we find
\begin{subequations}
\label{speq1pfg}
\begin{eqnarray}
\dot \phi_1 + \imath\, {\Phi} \phi_1 + \Bigl[\imath\, \frac{\bar X_0}{2V^{1/3 + 1/N_3}} + {\cG}\Bigr] \phi_3 &=& 0, \label{ph1}\\
\dot \phi_2 + \imath\, {\Phi} \phi_2 - \Bigl[\imath\, \frac{\bar X_0}{2V^{1/3 + 1/N_3}} - {\cG}\Bigr] \phi_4 &=& 0, \label{ph2}\\
\dot \phi_3 - \imath\, {\Phi} \phi_3 + \Bigl[\imath\, \frac{\bar X_0}{2V^{1/3 + 1/N_3}} -  {\cG}\Bigr] \phi_1 &=& 0, \label{ph3}\\
\dot \phi_4 - \imath\, {\Phi} \phi_4 - \Bigl[\imath\, \frac{\bar
X_0}{2V^{1/3 + 1/N_3}} + {\cG} \Bigr] \phi_2&=& 0. \label{ph4}
\end{eqnarray}
\end{subequations}
Further denoting ${\cal Y} = \frac{\bar X_0}{2V^{1/3 +1/ N_3}}$ we
can write the foregoing system of equation in the form:
\begin{equation}
\dot \phi = A \phi, \label{phi}
\end{equation}
with $\phi = {\rm
col}\left(\phi_1,\,\phi_2,\,\phi_3,\,\phi_4\right)$ and
\begin{equation}
A = \left(\begin{array}{cccc}-\imath\, \Phi &0 & -\imath\, {\cal Y}
- {\cG}& 0 \\ 0&-\imath\, \Phi& 0 &\imath\, {\cal Y} -
{\cG}\\-\imath\, {\cal Y} + {\cG}&0&\imath\, \Phi&0\\0&\imath\,
{\cal Y} + {\cG} & 0 &\imath\, \Phi \end{array}\right). \label{AMat}
\end{equation}
It can be easily found that
\begin{equation}
{\rm det} A = \left(\Phi^2 + {\cal Y}^2 +{\cG}^2\right)^2.
\label{detA}
\end{equation}

The solution to the equation \eqref{phi} can be written in the form
The solution to the equation \eqref{phi} can be written in the form
\begin{equation}
\phi(t) = {\rm T exp}\Bigl(-\int_t^{t_1}  A_1 (\tau) d \tau\Bigr)
\phi (t_1), \label{phi1}
\end{equation}
where
\begin{equation}
A_1 = \left(\begin{array}{cccc}-\imath\, {\cD} &0 & -\imath\, {\cal
Y} - {\cG}& 0 \\ 0&-\imath\, {\cD}& 0 &\imath\, {\cal Y} -
{\cG}\\-\imath\, {\cal Y} + {\cG}&0&\imath\, {\cD}&0\\0&\imath\,
{\cal Y} + {\cG} & 0 &\imath\, {\cD} \end{array}\right).
\label{AMat1}
\end{equation}
and $\phi (t_1)$ is the solution at $t = t_1$. As we have already
shown, $K = V_0^2/V^2$ for $K = \{J,\,I+J,\,I-J\}$ with trivial
spinor-mass and $K = V_0^2/V^2$ for $K=I$ for any spinor-mass. Since
our Universe is expanding, the quantities ${\cD}$, ${\cal Y}$ and
${\cG}$ become trivial at large $t$. Hence in case of  $K = I$ with
non-trivial spinor-mass one can assume $\phi (t_1) = {\rm
col}\left(e^{-\imath m_{\rm sp} t_1},\,e^{-\imath m_{\rm sp}
t_1},\,e^{\imath m_{\rm sp} t_1},\,e^{\imath m_{\rm sp}
t_1}\right)$, whereas for other cases with trivial spinor-mass we
have $\phi (t_1) = {\rm
col}\left(\phi_{1}^{0},\,\phi_{2}^{0},\,\phi_{3}^{0},\,\phi_{4}^{0}\right)$
with $\phi_i^0$ being some constants. Here we have used the fact
that $\Phi = m_{\rm sp} + {\cD}.$ The other way to solve the system
\eqref{speq1pfg} is given in \cite{saha2004b}.

As far as equation for $V$, i.e.,  \eqref{Vdefein} is concerned, we
solve it setting $K =  I$ as in this case we can use the mass term
as well. Assuming
\begin{equation}
F = \sum_{k} \lambda_k I^{n_k} =  \sum_{k} \lambda_k S^{2 n_k}
\label{nonlinearityIII}
\end{equation}
on account of $S = V_0/V$ we find
\begin{equation}
\ddot V = \Phi_1(V), \quad \Phi_1(V) = {\bar X} V^{\frac{1}{3} -
\frac{2}{N_3}} + \frac{3 \kappa}{2} \bigl[m_{\rm sp} V_0 +2 \sum_{k}
\lambda_k( 1 - n_k) V_0^{2n_k} V^{1 - 2n_k}\bigr].
\label{Vdefein1III}
\end{equation}
To determine the type of nonlinearity that can be dominant both at
the early stage as well as late time of evolution let us go back to
\eqref{Vdefein1III}. As one sees, for the nonlinearity to be
dominant at early stage when $V \to 0$ one should have $n_k = n_1$:
$n_1 > 1/2$ and $n_1 > 1/3 + 1/N_3$. For $n_k = n_0$: $n_0 = 1/2$
this term can be added to the mass term. And finally for the
nonlinear term to prevail at late time when $V \to \infty$ one
should choose $n_k = n_2$: $n_2 < 1/2$ and $n_2 < 1/3 + 1/N_3$. Then
we can rewrite the equation for $V$ with the nonlinear term that
determines both the early stage and the later stage of equation as
follows
\begin{eqnarray}
\ddot V &=& \Phi_1(V),  \label{Vdefein1IIInew} \\
\Phi_1(V) &=& {\bar X} V^{\frac{1}{3} - \frac{2}{N_3}} + \frac{3
\kappa}{2} \bigl[(m_{\rm sp} + \lambda_0) V_0 +2 \lambda_1( 1 - n_1)
V_0^{2n_1} V^{1 - 2n_1} + 2 \lambda_2( 1 - n_2) V_0^{2n_2} V^{1 -
2n_2}\bigr].\nonumber
\end{eqnarray}

with the first integral
\begin{eqnarray}
\dot V &=& \Phi_2(V), \label{1stintIII}\\
\Phi_2(V) &=& \sqrt{{\bar X_1} V^{(4N_3 - 6)/3N_3} + 3 \kappa
\left[(m_{\rm sp} + \lambda_0)V_0 V +  \lambda_1 V_0^{2n_1} V^{2(1 -
n_1)} +  \lambda_2 V_0^{2n_2} V^{2(1 - n_2)} \right] + {\bar C}
},\nonumber
\end{eqnarray}
where we denote $ {\bar X_1 } = 3 N_3 {\bar X}/(2N_3 - 3)$ and
${\bar C}$ is the constant of integration. The solution for $V$ can
be written in quadrature as

\begin{equation}
\int \frac{dV}{\Phi_2(V)} = t + t_0, \quad t_0 = {\rm const.}
\label{quadratureIII}
\end{equation}

In what follows we solve the Eqn. \eqref{Vdefein1III} numerically.
In doing so we determine $\dot V (0)$ from \eqref{1stintIII} for the
given value of $ V (0)$. To define whether the model allows
decelerated or accelerated mode of expansion we also study the
behavior of deceleration parameter $q$ defines as

\begin{equation}
q = - \frac{V \ddot V}{{\dot V}^2} = -\frac{V \Phi (V)}{\Phi_1^2
(V)}, \label{decelIII}
\end{equation}
which in view of \eqref{Vdefein1IIInew} and \eqref{1stintIII} reads

\begin{equation}
q = - \frac{{\bar X} V^{(4N_3 - 6)/3N_3} + \frac{3 \kappa}{2}
\bigl[(m_{\rm sp} + \lambda_0) V_0 V +2 \lambda_1( 1 - n_1)
V_0^{2n_1} V^{2(1 - n_1)} + 2 \lambda_2( 1 - n_2) V_0^{2n_2} V^{2(1
- n_2)}\bigr]}{{\bar X_1} V^{(4N_3 - 6)/3N_3} + 3 \kappa
\left[(m_{\rm sp} + \lambda_0)V_0 V +  \lambda_1 V_0^{2n_1} V^{2(1 -
n_1)} +  \lambda_2 V_0^{2n_2} V^{2(1 - n_2)} \right] + {\bar C}}.
\label{decelIIInew}
\end{equation}
From \eqref{decelIIInew} it can be easily established that

\begin{equation}
\lim_{V \to \infty}q  \longrightarrow - (1 - n_2) < 0, \quad {\rm
since} \quad n_2 < 1/2. \label{accele}
\end{equation}
Thus we see that spinor field nonlinearity generates late time
acceleration of the Universe.

Finally let us see, what happens to EoS (energy of state) parameter.
In view of \eqref{emt00}, \eqref{emtii} and \eqref{nonlinearityIII}
for the EoS parameter $\omega$ we find

\begin{equation}
\omega = \frac{p}{\ve} = - \frac{T_1^1}{T_0^0} = \frac{\sum_{k}
\lambda_k (2 n_k - 1) S^{2n_k}}{\sum_{k} \lambda_k S^{2n_k} + m_{\rm
sp} S}, \label{EoS}
\end{equation}
which on account of discussions above can be rewritten as

\begin{equation}
\omega =  \frac{\lambda_1 (2 n_1 - 1) V_0^{2n_1}V^{2n_2}  +
\lambda_2 (2 n_2 - 1) V_0^{2n_2}V^{2n_1} }{(\lambda_0 + m_{\rm sp})
V_0 V^{2(n_1+n_2) -1} + \lambda_1 V_0^{2n_1}V^{2n_2}  + \lambda_2
V_0^{2n_2}V^{2n_1} }. \label{EoS1}
\end{equation}

\begin{figure}[ht]
\centering
\includegraphics[height=70mm]{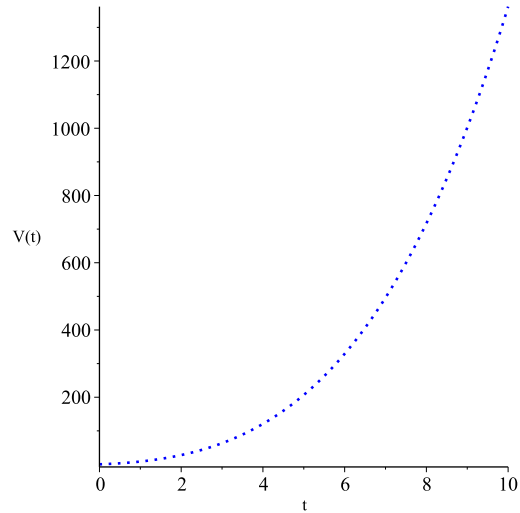} \\
\caption{Evolution of the Universe filled with massive spinor field
for a positive $\lambda_2$. Here we set $\lambda_1 =1$ and
$\lambda_2 = 1$} \label{Vbiiipos}.
\end{figure}

\begin{figure}[ht]
\centering
\includegraphics[height=70mm]{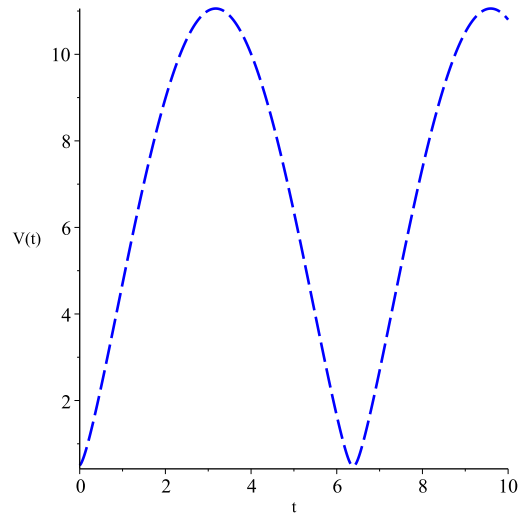} \\
\caption{Evolution of the Universe filled with massive spinor field
for a negative $\lambda_2$. In this case we set $\lambda_1 =-1$ and
$\lambda_2 = - 1$ } \label{Vbiiineg}.
\end{figure}

\begin{figure}[ht]
\centering
\includegraphics[height=70mm]{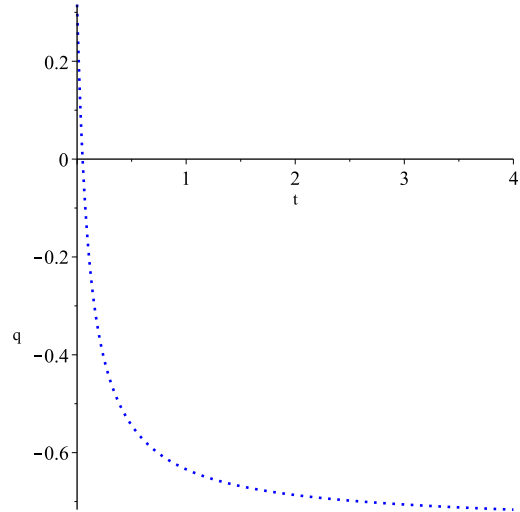} \\
\caption{Plot of deceleration parameter $q$ for a positive
$\lambda_2$. This case corresponds the evolution of volume scale
given in Fig. \ref{Vbiiipos}} \label{qbiiipos}.
\end{figure}

\begin{figure}[ht]
\centering
\includegraphics[height=70mm]{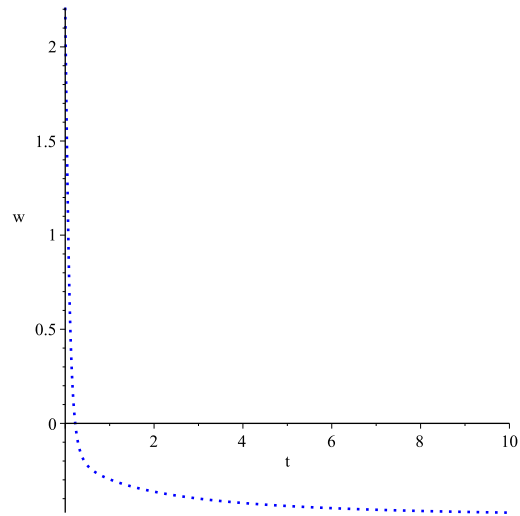} \\
\caption{Evolution of EoS parameter $w$ for a positive $\lambda_2$.
This case corresponds the evolution of volume scale given in Fig.
\ref{Vbiiipos}} \label{wbiiipos}.
\end{figure}

Since we are interested in qualitative picture here, so we set the
value of problem parameters very simple. Here we set $m = 1,\, X_1 =
1,\, X_0 = 1,\, V_0 = 1,\, \lambda_0 = 1,\, m_{\rm sp} = 1,\, C_0 =
1,\, \kappa = 1,\, N_3 =3.$ We consider two cases for different
combination with $\lambda_1 = \pm 1$ and $\lambda_2 = \pm 1$. It was
found that depending on the sign of $\lambda_2$ the model provides
two different type of solution, namely a positive $\lambda_2$ gives
rise to an expanding mode of evolution, whereas a negative
$\lambda_2$ generates oscillatory mode of evolution. In Figs.
\ref{Vbiiipos} and \ref{Vbiiineg} we plotted the evolution of the
Universe for a positive and negative value of $\lambda_2$,
respectively. The sign of $\lambda_1$ does not give a qualitatively
different picture. In Fig. \ref{qbiiipos} we have plotted the
dynamics of deceleration parameter $q$ that shows a late time
acceleration. In Fig. \ref{wbiiipos} we illustrated the EoS
parameter for a positive $\lambda_2$ that gives rise to an
accelerated mode of expansion. As one sees it is positive at the
beginning and becomes negative in the course of evolution which is
in correspondence with present day observations. It should be noted
that both deceleration and EoS parameters are time varying. This
fact is also in agreement with the modern picture of the evolution
of the Universe.

\section{Conclusion}

Within the scope of Bianchi type-III spacetime we study the role of
spinor field on the evolution of the Universe. Unlike Bianchi type
I, V or $VI_0$ cases where either both spinor mass and spinor field
nonlinearity vanish \cite{sahaIJTP2014,sahaAPSS2015,sahabvi0} or the
metric functions are similar to each other, i.e., $a_1 \sim a_2 \sim
a_3$ in this case no such problem occurs. As one can see from
\eqref{Metf} the spacetime remains locally rotationally symmetric
and anisotropic all the time, though the isotropy of the spacetime
can be achieved for a large proportionality constant. As far as
evolution is concerned, depending on the sign of coupling constant
the models allows both accelerated and oscillatory mode of
expansion. A negative coupling constat leads to an oscillatory mode
of expansion whereas a positive coupling constants generates
expanding Universe with late time acceleration. Both deceleration
parameter and EoS parameter in this case vary with time and are in
agreement with modern concept of spacetime evolution.

\vskip 0.1
cm

\noindent {\bf Acknowledgments}\\
This work is supported in part by a joint Romanian-LIT, JINR, Dubna
Research Project 4338-6-14/16, theme no. 05-6-1119-2014/2016. Taking
the opportunity I would also like to thank the reviewers for some
helpful discussions and references.

\end{document}